\documentclass{PoS}

\usepackage{amsmath}
\usepackage{slashed}
\usepackage{wasysym}
\usepackage{graphicx}
\usepackage{caption}
\usepackage{subcaption}
\newcommand{\RNum}[1]{\uppercase\expandafter{\romannumeral #1\relax}}

\title{Conformal Complex Scalar Singlet Extensions of the Standard Model: Symmetry Breaking Patterns and Phenomenology}

\ShortTitle{Conformal Complex Scalar Singlet Extensions of the Standard Model}

\author{\speaker{Zhi-Wei Wang}
       \\
        $\rm{CP}^3$-Origins, University of Southern Denmark \& University of Saskatchewan \& University of Waterloo\\
        E-mail: \email{wang@cp3.sdu.dk}}

\author{Frederick S. Sage%
      \\
      University of Saskatchewan\\
      E-mail: \email{frederick.sage@usask.ca}}

\author{T.G. Steele\\
        University of Saskatchewan\\
        E-mail: \email{tom.steele@usask.ca}}

\author{R.B. Mann\\
        University of Waterloo\\
        E-mail: \email{rbmann@uwaterloo.ca}}

\author{T.~Hanif\\
        University of Dhaka\\
        E-mail: \email{thanif@du.ac.bd}}

\abstract{We consider a conformal complex singlet extension of the Standard Model with a Higgs portal interaction. Two different scenarios depending on whether the global U(1) symmetry is broken or not have been studied. 
In the unbroken phase, the decay of the complex singlet is protected by the global U(1) symmetry which leads to an ideal cold dark matter candidate. In the broken phase, we are able to provide a second Higgs at $554\,\rm{GeV}$. In addition, gauging the global U(1) symmetry, we can construct an asymptotically safe U(1)' leptophobic model. We combine the notion of asymptotic safety with conformal symmetry and use the renormalization group equations as a bridge to connect UV boundary conditions and Electroweak (EW)/ TeV scale physics. We also provide a detailed example to show that these boundary conditions will lead to phenomenological signatures such as diboson excesses which could be tested at the LHC.

Preprint Number: $\rm{CP}^3$-Origins-2016-47 DNRF90}

\FullConference{38th International Conference on High Energy Physics\\
		  3-10 August 2016\\
		 Chicago, USA}

\begin{document}

\section{Motivation and Introduction}

Conformal models (CM) are particularly interesting since they may dynamically generate natural scale hierarchies through dimensional transmutation as in QCD in the Coleman-Weinberg framework \cite{Coleman:1973jx}. Moreover, there may exist a connection between the fixed point required in the (quantum) conformal scenario and the notion of asymptotic safety (AS) \cite{Wang:2015sxe}. Combining the notion of AS with conformal symmetry addresses the UV completion issue of the CM \cite{Wang:2015sxe}.

We study a conformal complex singlet extension of the SM with a Higgs portal interaction \cite{Wang:2015sxe,Wang:2015cda}. Two different scenarios depending on whether the global $U(1)$ symmetry is broken or not are discussed. In the broken phase, we predict a second Higgs at $554\,\rm{GeV}$ \cite{Wang:2015cda}. In the unbroken phase, the decay of the complex singlet is protected by the global $U(1)$ symmetry which leads to an ideal cold dark matter candidate \cite{Wang:2015cda}. In addition, gauging the global $U(1)$ symmetry, we can construct an asymptotically safe $U(1)'$ leptophobic model \cite{Wang:2015sxe}. We combine the notion of AS with conformal symmetry and use the renormalization group equations (RGE) as a bridge to connect UV boundary conditions and EW/ TeV scale physics. We also provide a detailed example to show the possible diboson excesses which could be detected at the LHC \cite{Wang:2015sxe}.

\section{Unbroken Phase}

The conformally symmetric complex singlet extension of the SM has the Lagrangian \cite{Wang:2015cda}:
\begin{equation}
\mathcal{L}=\partial_\mu H^{\dagger}\partial^\mu H+\partial_\mu S^{\dagger} \partial^\mu S-\lambda_2\left|S\right|^2 H^{\dagger}H-\lambda_3\left|S\right|^4-\lambda_1\left(H^{\dagger}H\right)^2\,.\label{Lagrangian}
\end{equation}
In the above, H is the (complex doublet) Higgs field, S is the complex singlet field and the Higgs portal interaction is proportional to $\lambda_2$. In this case $S$ decay is protected by the $U(1)$ global symmetry, making it an ideal cold dark matter candidate. Our analysis builds upon the Gildener-Weinberg method \cite{gildener} that generalizes the CW technique \cite{Coleman:1973jx} to incorporate multiple scalar fields. Letting $H=\frac{1}{\sqrt{2}}\left(\phi_1+i\phi_2, \phi_3+i\phi_4\right)$, $S=\frac{1}{\sqrt{2}}\left(\varphi_1+i\varphi_2\right)$ and defining $\phi^2=\sum_{i}\phi_i^2$ and $\varphi^2=\sum_{i}\varphi_i^2$, we obtain leading-logarithm expression for the effective potential
\cite{Wang:2015cda}
\begin{equation}
V_{LL}=\frac{1}{4}\lambda _1 \phi^4+\frac{1}{4}\lambda _2\phi^2\varphi^2+\frac{1}{4}\lambda _3\varphi^4+BL+CL^2+DL^3+EL^4+\ldots\label{VLL}
\end{equation}
where $L\equiv\log\left(\frac{\phi^2+\varphi^2}{\mu^2}\right)$. 
The quantities $B, C, D, E$ are the functions of $\left(\lambda_1,\lambda_2,\lambda_3,g_t,\phi,\varphi\right)$ which are dimension-4 combinations of $\phi^2$ and $\varphi^2$ as required by 
 symmetry and
contain leading-logarithm ($LL$) combinations of couplings.
The coefficients $B, C, D, E$ are determined by RGE
\begin{equation}
\left(\mu\frac{\partial}{\partial\mu}+\beta_{g_t}\frac{\partial}{\partial g_t}+\sum_{i=1}^3\beta_i\frac{\partial}{\partial \lambda_i}+
\gamma_\phi \phi\frac{\partial}{\partial \phi}\right)V_{LL}=0\,,\label{rg equation}
\end{equation}
where $\beta_i,\beta_{g_t}$ and $\gamma_\phi$ are the one loop RG functions and and anomalous dimensions respectively.

Truncation of the effective potential at $LL$ order leads to
\begin{equation}
V_{eff}=V_{LL}+K_1\phi^4+K_2\phi^2\varphi^2+K_3\varphi^4 \,,
\label{counter1}
\end{equation}
where counter terms $K_i$ are functions of the couplings and can be determined by renormalization conditions \cite{Coleman:1973jx}\cite{Wang:2015cda}. 
The couplings and the masses of $M_H$ and singlet $M_S$ are given resepctively by
\begin{equation}
\frac{dV_{eff}}{d\phi}\bigg|_{\phi=v\atop \varphi=v_1}=0\,,\frac{dV_{eff}}{d\varphi}\bigg|_{\phi=v\atop \varphi=v_1}=0\quad;\quad M_{H}^2=\frac{dV_{eff}^2}{d\phi^2}\big|_{\phi=v\atop \varphi=v_1},\quad M_{S}^2=\frac{dV_{eff}^2}{d\varphi^2}\big|_{\phi=v\atop \varphi=v_1}\,,\label{constraint2}
\end{equation}
where $v$ is identified with the EW scale $v=246.2\,\rm{GeV}$ and the singlet scale $v_1=0$ corresponding to the unbroken case.

In Fig.~\ref{lux}, we illustrate our predicted dark matter mass/coupling relation in the green curve. The abundance curves (orange and blue) are calculated using the results of Ref.~\cite{Cline:ab}. Setting the dark matter self-interaction coupling to $\lambda_3=1$ shifts the results slightly from the green to the purple curve in the figure, retaining this qualitative feature. The shaded region in Fig.~\ref{lux} represents the parameter space excluded by the LUX experiment at $95\%$ CL \cite{Akerib:2013tjd}. Most of the parameter space below $85\,\rm{GeV}$ is ruled out by the LUX experiment \cite{Akerib:2013tjd}, apart from a small region of parameter space in the $M_S\approx M_{H}/2$ resonant region, which is strongly constrained by the Higgs decay width \cite{Cline:ab}. 
Combining the LUX \cite{Akerib:2013tjd} and dark matter abundance constraints, the complex singlet model admits a viable dark matter candidate $100\rm{GeV}\leq M_s\leq110\rm{GeV}$ with Higgs portal interaction $0.05\leq\lambda_2\leq0.2$ corresponding to $10\%-100\%$ dark matter abundance. 
The viable dark matter candidates resulting from our analysis are very close to the boundary of the current direct detection experiments and will be in the detection region of the coming experiments XENON1T \cite{XENON1T} results. 



\section{Broken Phase (non-gauged $U(1)$) and gauged $U(1)$}

The broken-symmetry case $\langle S\rangle\neq0$  is particularly interesting since the real component of the complex singlet will mix with the SM Higgs field, leading to one heavy and one light Higgs field. The light state corresponds to the $125\,\rm{GeV}$ observed Higgs boson.  
By using the generalized optimization method developed in \cite{Wang:2015cda}, we have four constraints for five parameters $\lambda_1(t^*)$, $\lambda_2(t^*), \lambda_3(t^*), v_1, t^*$ where $\langle S\rangle=v_1$ is the VEV of the singlet field and $t^*$ is the optimized scale. Using the $125\,\rm{GeV}$ Higgs mass as an extra constraint, we find 
an additional heavy Higgs at $554\,\rm{GeV}$.

Gauging the $U(1)$ symmetry, the derivative in Eq.~\eqref{Lagrangian} will be replaced by covariant derivative $D_\mu$.
In the basis where the two $U(1)$ gauge kinetic terms are diagonal, $D_\mu$ is written as \cite{Wang:2015sxe}
\begin{equation}
D_\mu=\partial_\mu-ig_3\frac{\lambda_a}{2}G_\mu^a-ig_2\frac{\tau_i}{2}W_\mu^i-iY\left(g_YB_\mu+g_{m}B_\mu'\right)-ig'Q_B'B_\mu',
\end{equation}
where $g_3$, $g_2$, $g_Y$ and $g'$ are the gauge couplings of $SU(3)_c$, $SU(2)$, $U(1)_Y$ and $U(1)'$ respectively. The quantities $Y$ and $Q_B'$ denote the $U(1)_Y$ hypercharge and the $U(1)'$ charge.  We make explicit the mixing term proportional to $g_m$ that couples the $B_\mu^\prime$ field to SM hypercharge $Y$. Dilepton constraints on new neutral gauge bosons are stringent, so we would like to avoid coupling the $U(1)^\prime$ gauge group to SM leptons, making the model leptophobic.We choose a special case of the gauge group $U(1)_{B-xL}'$ where $x=0$ and the gauge group in our case can be denoted $U(1)_B'$ with charge $Q_B'$ \cite{Carena:2004xs}. The charge assignments of the $U(1)_B'$ model are summarized in Table \ref{fermion gauge charge} where $\nu_R$ and two `spectator' fermions $\psi^l_L$ and $\psi^e_L$ are introduced to cancel the anomaly.

\begin{figure}[htb]
\centering
\begin{subfigure}{0.4\textwidth}
\centering
\includegraphics[width=\linewidth]{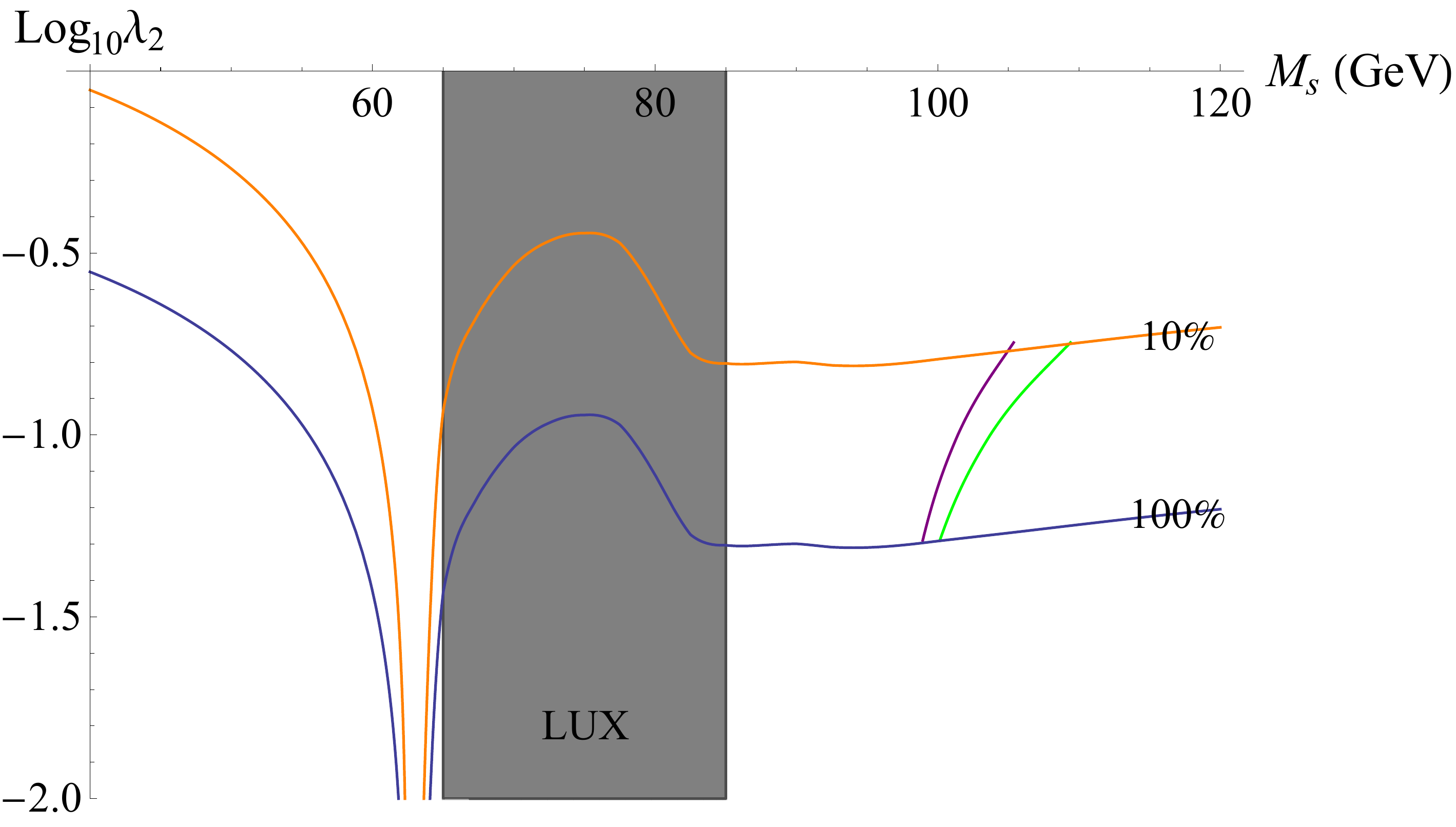}
\caption{Relationship between predicted dark matter mass and Higgs portal coupling}
\label{lux}
\end{subfigure}
\begin{subfigure}{0.4\textwidth}
\centering
\includegraphics[width=\linewidth]{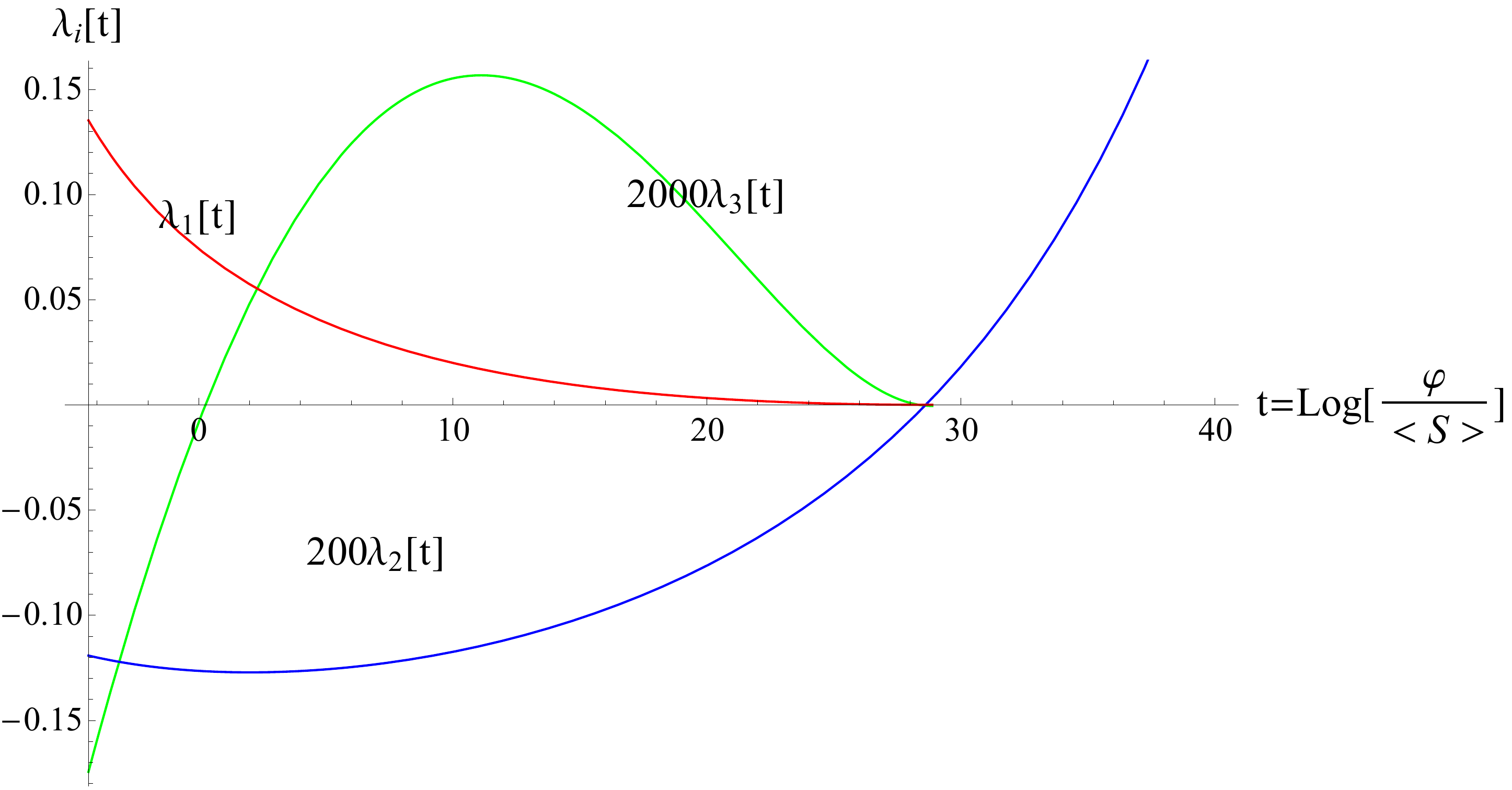}
\caption{Running scalar couplings are shown as a function of the scale $t=\log\left(\varphi/\langle S\rangle\right)$. The red, blue, and green curves represent $\lambda_1\left(t\right)$, $200\lambda_2\left(t\right)$, $2000\lambda_3\left(t\right)$ respectively. }
\label{running coupling3}
\end{subfigure}
\end{figure}

\begin{table}[ht]
\centering
\begin{tabular}{||c||c|c|c|c|c|c|c|c|c||}
    	\hline
Fermion & $q_L$ & $u_R$ & $d_R$ & $l_L$ & $e_R$ & $\nu_R$ & $\psi_L^l$ & $\psi_L^e$ & $\chi$ \\ \hline $U\left(1\right)^\prime$ Charge & 1/3 & 1/3 & 1/3 & 0 & 0 & -1 & -1 & -1 & 1 \\ \hline
\end{tabular}
\caption{Fermion gauge charges.}
\label{fermion gauge charge}
\end{table}
We provide a categorization of different AS scenarios according to the gravity contribution to the RG functions above Planck scale \cite{Wang:2015sxe}: 
\begin{align}
\beta_{\lambda_1}\left(\Lambda\right)&=\beta_{\lambda_3}\left(\Lambda\right)=\lambda_1\left(\Lambda\right)=\lambda_2\left(\Lambda\right)=\lambda_3\left(\Lambda\right)=0\label{boundary1}\\
\beta_{\lambda_1}\left(\Lambda\right)&=\lambda_1\left(\Lambda\right)=0;~ \lambda_2\left(\Lambda\right),\, \lambda_3\left(\Lambda\right)\neq0\label{boundary2}\\
\beta_{\lambda_1}\left(\Lambda\right)&=\lambda_1\left(\Lambda\right)=\lambda_2\left(\Lambda\right)=0,~ \lambda_3\left(\Lambda\right)\neq0\label{boundary3}\,.
\end{align}
We use the RG equation as a bridge to connect the UV boundary conditions to EW/TeV scale physics and explore implications for SM observables. The predictive power of AS scenarios implies that most of the parameters in the model are uniquely determined, thereby providing interesting interrelationships among the couplings,  the scale of the fixed point (transition scale) $M_{UV}$, and the generations of quarks coupled to the $U(1)^\prime$ gauge field \cite{Wang:2015sxe}. We plot the running scalar couplings from the EW scale to the UV transition scale in Figure \ref{running coupling3} by using the boundary condition \eqref{boundary1}.

We are able to show that assuming the $Z'$ mass is around $2\,\rm{TeV}$, the boundary conditions \eqref{boundary1}--\eqref{boundary3} would lead to diboson excesses $Z'\rightarrow ZZ$ at $0.03\,\rm{fb}$, $0.23-4\,\rm{fb}$ and $0.01\,\rm{fb}$ respectively.



\end{document}